\documentclass[11pt]{article}
\usepackage{setspace}
\usepackage{graphicx}
\usepackage{lscape}
\usepackage{pdflscape}
\usepackage{afterpage}
\usepackage{pdfpages}
\usepackage{float}
\usepackage{booktabs}
\usepackage[utf8]{inputenc}
\usepackage{indentfirst}
\usepackage{array}
\usepackage{arydshln}
\usepackage{arydshln}
\usepackage{ragged2e}
\usepackage{tikz}
\usepackage{lipsum}
\usepackage{pgffor}
\usepackage{dsfont}
\usepackage{dsfont}
\usepackage{amsfonts}
\usepackage{amsmath}
\allowdisplaybreaks
\usepackage{mathtools}

\usepackage{xfrac}
\usepackage{amssymb}
\usepackage{bigints}
\usepackage{graphicx}
\usepackage{url}				
\usepackage{caption}
\usepackage{subcaption} 
\usepackage{enumitem}
\usepackage{relsize,exscale}
\usepackage{multirow}
\usepackage{natbib}
\setcitestyle{authoryear, open={(},close={)}}
\usepackage{datetime}
\usepackage{comment}
\usepackage{array}
\usepackage{pifont}
\newcommand{\xmark}{\ding{55}}%

\usepackage[flushleft]{threeparttable}
\usepackage{siunitx}
\usepackage{bbm}

\newdateformat{monthyeardate}{%
	\monthname[\THEMONTH] \THEYEAR}

\usepackage{titlesec}
\titleformat*{\section}{\large\bfseries}
\titleformat*{\subsection}{\large\bfseries}


\newcounter{parentnumber}
\usepackage{theorem}

\newtheorem{assumption}{Assumption}

\newtheorem{proposition}{Proposition}

\usepackage[letterpaper, margin=1in]{geometry}

\usepackage{hyperref} 
\hypersetup{
	colorlinks=true, breaklinks=true, bookmarks=true,bookmarksnumbered,
	urlcolor=blue, linkcolor=blue, citecolor=blue, 
	pdftitle={}, 
	pdfauthor={\textcopyright}, 
	pdfsubject={}, 
	pdfkeywords={}, 
	pdfcreator={pdfLaTeX}, 
	pdfproducer={LaTeX with hyperref and ClassicThesis} 
}

\begin{document}
	\setstretch{1}
	\title{{\LARGE Beyond Linearity: Semiparametric Solutions to Contamination Bias with Multi-valued Treatments\thanks{We would like to thank Pedro Sant'Anna and Gabriela Szini for helpful suggestions. A replication package is available at \url{https://github.com/Micsifin/Contamination-Bias-Semiparametric-Solutions}.}}}

	\author{
	Michel Csillag Finger\thanks{São Paulo School of Economics - FGV and Institute of Mathematics and Statistics - USP.  Email: \href{mailto:michel.finger@usp.br}{michel.finger@usp.br}.}  \and Vitor Possebom\thanks{São Paulo School of Economics - FGV. Email: \href{mailto:vitor.possebom@fgv.br}{vitor.possebom@fgv.br}. This study was financed, in part, by the São Paulo Research Foundation (FAPESP), Brazil. Process Number \#2025/04857-0.}}
	\date{}

	\maketitle

	\newsavebox{\tablebox} \newlength{\tableboxwidth}


\begin{center}

First Draft: September 2026; This Draft: \monthyeardate\today


%
%
\href{https://sites.google.com/site/vitorapossebom/working-papers}{Please click here for the most recent version}

\

		\large{\textbf{Abstract}}
	\end{center}

    We examine semiparametric solutions to contamination bias for nonbinary treatments. Deepening the discussion by \cite{gphk2024contamination}, we detail how spline functions approximate conditional expectation and propensity score functions under weak functional-form assumptions. Reanalyzing 15 regressions across 11 studies from top Economics journals, we compare standard linear regressions against parametric and semiparametric versions of three contamination-robust estimators. We document large discrepancies between parametric and semiparametric point-estimates and find that adopting flexible semiparametric solutions may not increase statistical uncertainty. We recommend that researchers verify their conclusions' robustness to the use of semiparametric tools that address contamination bias.

	\


        \


	\newpage

	\doublespacing

\section{Introduction}\label{sec:Intro}

Multi-valued discrete treatments are widespread in observational studies and randomized controlled trials. For example, \cite{weisburst2019police} focuses on police racial bias in use of force and considers four racial categories: White, Black, Hispanic, and other. \cite{brownback2020instruction} study the impact of four types of performance-based incentives for community college instructors and students. These settings are typically analyzed with regressions of an outcome variable on indicator variables for each treatment arm and control variables. However, \citet{gphk2024contamination} show that these regression coefficients may present contamination bias, i.e., they are functions of the average effects of the treatment of interest and all other treatment arms. To avoid these issues, they propose three identification strategies that avoid contamination bias. Although they briefly discuss nonparametric implementation, they focus their estimation and empirical exercises on linear parametric estimators.

Because these estimators rely on strong linearity assumptions, we deepen their discussion about avoiding contamination bias through the usage of semiparametric smoothing functions. In particular, we detail how to use spline functions to approximate the necessary propensity score and conditional expectation functions. By doing so, we are able to easily implement semiparametric versions of the three estimators proposed by \cite{gphk2024contamination} using the statistical package provided by \cite{multe2024}. Since these semiparametric versions rely on weaker functional-form restrictions, they increase the credibility of the three solutions proposed by \cite{gphk2024contamination} for addressing contamination bias.

Next, we compare our proposed spline estimators against the parametric solutions implemented by \cite{gphk2024contamination} and against the usual linear regression estimator in relevant empirical settings. To do so, we replicate articles from top Economics journals. The replicated articles include six studies with continuous covariates previously analyzed by \citet{gphk2024contamination}, plus five additional articles from American Economic Review, Econometrica, and the Journal of Political Economy. As a result, we substantially increase the number of regressions used to compare seven estimators: standard linear regression with a full set of treatment-arm indicators, the parametric versions of the three solutions proposed by \cite{gphk2024contamination}, and our spline versions of these three solutions. Specifically, we analyze 15 regressions from 11 articles, for a total of 141 parametric estimates and 141 semiparametric estimates.

We present three sets of results concerning contamination bias in standard linear regressions, differences in the point estimates of parametric and semiparametric approaches, and differences in the statistical uncertainty behind the seven analyzed estimators. First, similarly to \cite{gphk2024contamination}, we document large variations in point-estimates when we compare the standard linear regression estimator against the strategies proposed by \cite{gphk2024contamination}. These discrepancies suggest that contamination bias might be a relevant problem in many empirical settings with nonbinary treatments. Second, we find substantial differences between the point estimates of the linear parametric estimators implemented by \cite{gphk2024contamination} and the semiparametric spline estimators discussed in this article. Third, when we compare the estimated standard errors of the seven analyzed estimators, we find that adopting approaches that address contamination bias may not increase statistical uncertainty even when these approaches rely on flexible semiparametric estimators.

We contribute to two strands of the economics literature. First, we extend the discussion about how to account for treatment effect heterogeneity. The mismatch between nonparametric, fully heterogeneous identification strategies and parametric estimators has been discussed by \cite{GoodmanBacon2021, dechaisemartin2020twowayFE, sun2021dynamic, callawaysantanna2021did} in the differences-in-differences setting, \cite{Sloczynski2022} in the binary treatment context, \cite{słoczyński2024} in the instrumental variable framework, \cite{Gelman2019} in the regression discontinuity setting, and \cite{gphk2024contamination} in the nonbinary treatment context. We contribute especially to the last setting by discussing in detail the semiparametric solutions suggested by \cite{gphk2024contamination}.

Second, we contribute to the empirical practice literature. Understanding how different estimators behave in relevant empirical settings has been a fruitful area of research. In instrumental variable contexts, \cite{Lal2024} discuss different ways to measure statistical uncertainty and instrument strength. In difference-in-differences settings, \cite{Bertrand2004} compare different approaches to measuring statistical uncertainty, while \citet{chiu2025analysis} discuss different estimators that account for heterogeneous effects and staggered treatment adoption. In the multi-valued discrete treatment setting, \cite{gphk2024contamination} compare three parametric solutions to contamination bias using nine standard regressions, for a total of 90 parametric estimates. We contribute especially to this last setting by analyzing semiparametric versions of these estimators and increasing the number of regressions analyzed to 15, for a total of 282 estimates.

The paper is organized as follows. Section \ref{sec:theory} explains the contamination bias problem in standard linear regression with multi-valued discrete treatment and presents three parametric and three semiparametric solutions to this issue. Section \ref{sec:method} explains how we selected the replicated articles and regressions used to assess the relevance of semiparametric estimators in a setting with nonbinary treatments. Section \ref{sec:results} discusses our main empirical findings, while Section \ref{sec:conclusion} concludes.




\section{Theoretical Background}\label{sec:theory}

We explain the identification issues related to contamination bias in Section \ref{sec:bias} and discuss six solutions in Section \ref{sec:solutions}. Although \citet{gphk2024contamination} already discussed these solutions, we contribute by analyzing semiparametric solutions in detail. We keep the exposition succinct for brevity.

\subsection{Understanding Contamination Bias} \label{sec:bias}

Suppose a linear regression model with an outcome variable $Y_i$, a $\left(K+1\right)$-dimensional treatment vector $X_i$, and covariates $W_i$. Specifically, each component $k \in \left\lbrace 1, \ldots, K \right\rbrace$ of $X_i$ is a binary variable $X_{k,i}$ indicating that individual $i$ received treatment arm $k$, where all $K$ treatment arms are mutually exclusive and $k = 0$ denotes the omitted control group with $X_{0,i} = 1-\sum_{k=1}^{K}X_{k,i}$. The covariates enter the regression through a function $g \in \mathcal{G}$, where $\mathcal{G}$ is a linear space of functions. Mathematically, this regression model is represented by
\begin{equation}
\label{eq:gen-model}
Y_i = X_i^{\prime} \beta + g(W_i) + U_i,
\end{equation}
where $U_i$ is the residual and $\beta$ and $g$ are defined by
\begin{equation}
\label{eq:gen-b-g}
(\beta, g) \coloneqq \mathop{\arg \min}_{\tilde{\beta} \in \mathbb{R}^K, \tilde{g} \in \mathcal{G}} \mathbb{E}[(Y_i - X_i^{\prime} \tilde{\beta} -\tilde{g}(W_i))^2].
\end{equation}

Let $Y_i (k)$ denote the potential outcome of unit $i$ under treatment $k$. Define the vector of treatment effects $\tau_i$ with elements $\tau_{ik} = Y_i (k) - Y_i (0)$. The vector $\tau (W_i) = \mathbb{E} [\tau_i | W_i]$ captures the conditional average treatment effects (CATE), where $\tau_k(W_i)$ corresponds to the CATE of treatment $k$. Define also the vector of propensity scores $p(W_i) = \mathbb{E}[X_i | W_i]$, where $p_k (W_i) = \mathbb{P} \left[X_{k,i} = 1 | W_i\right]$ is the conditional probability of receiving treatment arm $k$. 

Following \citet{gphk2024contamination}, we make two assumptions:
\begin{assumption} \label{ass:mean-independence}
Given covariates, potential outcomes are mean-independent of the treatment, i.e., $\mathbb{E}[Y_i (k) | X_i, W_i] = \mathbb{E}[Y_i (k) | W_i]$ for all $k$.
\end{assumption}

\begin{assumption} \label{ass:g-specified}
The linear space $\mathcal{G}$ is flexible enough to contain either the true propensity score functions or the true conditional expectation of the untreated potential outcome. Formally, we have that either $p_k \in \mathcal{G}$ for all $k$ or $\mu_0 \in \mathcal{G}$, where $\mu_0 (w) := \mathbb{E}[Y_i (0) | W_i = w]$.
\end{assumption}

\citet[Proposition 1]{gphk2024contamination} shows that the regression coefficient $\beta_{k}$ of Equation \eqref{eq:gen-model} is a sum of two terms, where the first term is a convex combination of the $k$-th treatment arm's CATEs and the second term is a possibly non-zero linear combination of the CATEs of the other treatment arms. Formally, \citet[Proposition 1]{gphk2024contamination} derives the following result:
\begin{proposition} \label{prop:general-bias}
Under Assumptions \ref{ass:mean-independence} and \ref{ass:g-specified}, the treatment coefficients in Equation \eqref{eq:gen-model} identify
\begin{equation} \label{eq:general-bias}
\beta_k = \mathbb{E}[\lambda_{kk}(W_i) \tau_{k}(W_i)] + \sum_{\ell \neq k} \mathbb{E}[\lambda_{k\ell}(W_i) \tau_{\ell}(W_i)],    
\end{equation}
where $\mathbb{E}[\lambda_{kk}(W_i)] = 1$ and $\mathbb{E}[\lambda_{k\ell}(W_i)] = 0$. Moreover, contamination weights $\lambda_{k\ell}(W_i)$ will be non-zero unless treatment assignment is unconditionally random (i.e., $p_k (W_i) = \mathbb{P} \left[X_{k,i} = 1 | W_i\right]$ is a constant function for every $k$).
\end{proposition}

In settings where treatment assignment depends on covariates, Proposition \ref{prop:general-bias} implies that the coefficient $\beta_k$ does not identify any sort of average of the corresponding $k$-th treatment arm's CATEs. It shows that this regression coefficient identifies a linear combination of \textit{all} CATEs due to the contamination bias term in Equation \eqref{eq:general-bias}.


\subsection{Avoiding Contamination Bias} \label{sec:solutions}

\citet{gphk2024contamination} propose three solutions to identify interpretable convex combinations of CATEs in settings with multi-valued discrete treatments. As pointed out in \citet[p.~4031]{gphk2024contamination}, a key assumption in all these strategies is that the conditional expectation function of each potential treatment outcome is in the space of functions $\mathcal{G}$. For this reason, \citet[e.g., Equation (16)]{gphk2024contamination} suggest using nonparametric smoothing functions (e.g., splines) to approximate the relevant conditional expectation function and propensity scores. Despite this suggestion, in their empirical analysis, \citet[Sections III.C and IV]{gphk2024contamination} implement only linear estimators that restrict $\mathcal{G}$ to be the space of linear functions of the covariates.

Our main contribution is to follow their suggestion of using nonlinear methods to approximate the relevant conditional expectation functions and propensity scores. We empirically implement their three solutions using splines as a semiparametric approximation to these functions. Beyond reanalyzing all the articles that were discussed by \cite{gphk2024contamination} and contain continuous covariates, we include additional articles published in the Journal of Political Economy, Econometrica, and the American Economic Review. As a consequence, we substantially increase the number of regressions that are used to compare seven estimators: standard linear regression (Equation \eqref{eq:simple-reg}), the parametric versions of the three solutions proposed by \cite{gphk2024contamination} and the spline version of these three solutions.

To understand these seven estimators, we need to build our notation. Define the expected conditional potential outcome as $\mu_k (w) = \mathbb{E}[Y_i (k) | W_i = w]$ for $k = 0, \dots, K$, implying that $\tau_k (w) = \mu_k (w) - \mu_0 (w)$. Define a transformation of the covariates $q_k (w)$ such that $q_k (\cdot) \in \mathcal{G}$. For example, we can choose $\mathcal{G}$ to contain infinite sequences of semiparametric smooth basis functions $\{ b_j (W_i) \}_{j = 0}^{\infty}$.

As previously mentioned, we focus on splines as these semiparametric smooth basis functions. Specifically, we use the definition of univariate splines by \citet[p.~5571]{chen2007}: for a spline of order $r$ with $J$ knots for a continuous variable $W_{i}$, we use
\begin{equation} \label{eq:spline-def}
\text{Spl}(r, J; W_i) = \sum_{\ell=1}^{r-1} W_i^\ell + \sum_{j=1}^{J} [\max \{ W_i - t_j, 0 \} ]^{r-1},
\end{equation}
where the knots are given by $t_j \text{ with } j = 0, \dots, J + 1$ such that $\min \{ W_i \} = t_0 < t_1 < t_2 < \dots < t_{J-1} < t_{J} < t_{J+1} = \max \{ W_i \}$.

It is important to note that the spline definition in Equation \eqref{eq:spline-def} only applies for continuous covariates.
To include categorical and binary covariates, we use a full set of indicator variables. Specifically, we define $S(W_i) \in \mathcal{G}$ such that
\begin{equation} \label{eq:qk-spline}
S (W_i) = 
\begin{cases}
\text{Spl} (r, J; W_{ij}), & \text{if } W_{ij} \text{ is a continuous variable} \\
W_{ij}, & \text{if } W_{ij} \text{ is a binary variable} \\
\sum_{\ell = 2}^{L}\mathbbm{1}\{ W_{ij} = \ell \} , & \text{if } W_{ij} \text{ is a categorical variable, with } L \text{ categories}
\end{cases}
\end{equation}
as one possible transformation of the covariates, which we will use in the next sections. Note that this transformation imposes additive separability between the covariates, which could be avoided by exploring tensor splines.

In the next three sections, we present the three solutions proposed by \citet{gphk2024contamination} with their respective spline extensions.
Each solution should be compared to a ``simple'' regression, which is the ``canonical'' method of estimating $\hat{\beta}$, with a coefficient for each treatment and for each covariate.
It is equivalent to choose $\mathcal{G} = \{\alpha + w'\gamma : [\alpha, \gamma']' \in \mathbb{R}^{1+\text{dim}(W_i)}\}$.
Explicitly, this model may suffer from contamination bias and is given by Equation \eqref{eq:simple-reg}.

\begin{equation} \label{eq:simple-reg}
    Y_i = \alpha_{0} + \sum_{k=1}^{K} X_{k,i} \beta_k + W_i' \gamma + U_i
\end{equation}

\subsubsection{Estimating Average Treatment Effects (ATE)}

\citet{gphk2024contamination} extend the ATE estimator proposed for binary treatments by \citet[Subsection~5.3]{imbens2009developments} and use the following regression model:
\begin{equation} \label{eq:ate}
    Y_i = \dot{\alpha}_0 + \sum_{k = 1}^{K} X_{k,i} \dot{\beta}_k + W_i' \alpha_{W,0} + \sum_{k = 1}^{K} X_{k,i} (W_i - \overline{W})' \gamma_{W, k} + \dot{U}_i
\end{equation}
where $\overline{W} = \frac{1}{N} \Sigma_{i=1}^{N} W_i$ and $N$ is the number of observations. There are two main differences from the standard regression in Equation \eqref{eq:simple-reg}. First, we add interaction terms to ensure each treatment coefficient depends only on the outcome of interest and the untreated outcome, avoiding contamination bias if the conditional expectation functions of the potential outcomes are linear in the covariates. Second, covariates $W_i$ are demeaned to ensure that the coefficients appropriately identify the ATE if the conditional expectation functions are linear in the covariates.

If the relevant conditional expectation functions are not linear, we follow the suggestion by \citet{gphk2024contamination} and implement a regression specification with splines:
\begin{equation} \label{eq:ate-spline}
Y_i = \dot{\alpha}_{S,0} + \sum_{k = 1}^{K} X_{k,i} \dot{\beta}_{S,k} + S (W_i)^{\prime} \alpha_{S, 0} + \sum_{k=1}^{K} X_{k,i} \left(S (W_i) - \overline{S} (W_i) \right)^{\prime} \gamma_{S, k} + \dot{U}_{S,i}
\end{equation}
where $\overline{S} (W_i) = \frac{1}{N} \sum_{i=1}^{N} S (W_i)$. Note that this spline estimator achieves the semiparametric efficiency bound under the strong overlap condition \citep{chen2007} and that its standard errors can be estimated using the usual parametric formulas \citep{Ackerberg2012}.

\subsubsection{Easiest-to-Estimate Averages of Treatment Effects (EW)}\label{SecEW}

Instead of interacting the covariates with the treatment-arm indicator variables, \citet[Equation (24)]{gphk2024contamination} also propose running separate regressions for each treatment arm, i.e.,
\begin{equation}
\label{eq:ew-model} Y_i = \ddot{\alpha}_k + X_{k,i} \ddot{\beta}_k + W_i' \ddot{\gamma}_k + \ddot{U}_{ik},
\end{equation}
where we use a sample containing only units assigned to either treatment arm $k$ or the control group. According to \citet{gphk2024contamination}, coefficient $\ddot{\beta}_k$ captures a weighted average of treatment arm $k$'s CATEs if the conditional expectation functions of the potential outcomes are linear in the covariates.

If the relevant conditional expectation functions are not linear, we follow the suggestion by \citet{gphk2024contamination} and implement a regression specification with splines:
\begin{equation}
\label{eq:ew-spline} Y_i = \ddot{\alpha}_{S,k} + X_{k,i} \ddot{\beta}_{S,k} + S (W_i)' \ddot{\gamma}_{S,k} + \ddot{U}_{S,ik},
\end{equation}
where we use a sample containing only units assigned to either treatment arm $k$ or the control group.

\subsubsection{Easiest-to-Estimate Common Weighting (CW) Scheme}\label{SecCW}

Although the EW solution in Section \ref{SecEW} is easy to estimate and has interesting efficiency properties according to \citet[Corollary 1]{gphk2024contamination}, it only allows comparisons between treatment $k$ and the control group. To facilitate the comparison between any treatment arms while putting more weight on covariate values that satisfy the overlap condition, \cite{gphk2024contamination} propose a weighted least squares estimator that uses the full sample and regresses the outcome variable $Y_i$ on treatment vector $X_i$ and a constant, where unit $i$'s weight is given by
\begin{equation}
    \label{eq:cw}
    \left( \sum_{k=0}^{K} \frac{\pi_k (1 - \pi_k)}{p_k (W_i)} \right)^{-1} \cdot \left(p_{D_{i}} (W_i)\right)^{-1},
\end{equation}
where $\pi_k$ is the unconditional probability of treatment arm $k$, and $D_{i} = k$ if and only if $X_{k,i} = 1$. To feasibly implement this regression, the unit weights use $\hat{\pi}_k = \sfrac{\sum_{i=1}^{N} X_{k,i}}{N}$ for $\pi_k$ and parametric multinomial logit estimators $\hat{p}_k (W_i)$ for the propensity score functions $p_k (W_i)$. The k-coefficient in this weighted regression captures a weighted average of treatment arm $k$'s CATEs if the linearity assumptions behind the propensity score estimators hold.

If these linearity assumptions do not hold, we follow the suggestion by \citet{gphk2024contamination} and implement a regression specification with splines for the propensity score estimators. To do so, we apply the spline transformation on the covariates and use these transformed variables in the multinomial logit estimator.

\section{Comparing Solutions to Contamination Bias in Empirical Settings}\label{sec:method}

To understand the importance of semiparametric solutions to contamination bias, it is paramount to compare our proposed spline estimators against the parametric solutions implemented by \cite{gphk2024contamination} and against the usual linear regression estimator (Equation \eqref{eq:simple-reg}) in relevant empirical settings. To do so, we replicate articles from top Economics journals. The replicated articles include six studies with continuous covariates previously analyzed by \citet{gphk2024contamination}, plus five additional articles from American Economic Review, Econometrica, and the Journal of Political Economy.

To select the articles in our replication exercise, we follow closely the method used by \citet[Appendix C.1]{gphk2024contamination}. However, these authors included only articles published by the American Economic Association while we also searched for articles published in Econometrica, the Journal of Political Economy, the Quarterly Journal of Economics, and the Review of Economic Studies. Specifically, articles from these journals were filtered for containing in the title, abstract, or body any word related to ``experiments'' (\textit{stratified, random, RCT, experiment}) or to racial disparities (\textit{racial/ethnic differences, discrimination, disparities, gaps}). The resulting articles were manually analyzed to verify (i) the presence of multi-valued discrete treatments, and (ii) the public availability of a replication package.

Using this procedure, we found eight experimental articles and three observational studies. Table \ref{tab:articles} lists those articles and the replicated columns of each of their published tables. Unlike \cite{gphk2024contamination}, we reanalyze more than one regression per article, implying that we analyze a total of 15 standard regressions, 141 parametric estimates, and 141 semiparametric estimates. In comparison, \cite{gphk2024contamination} analyze 9 standard regressions, 30 treatment effects, and 90 parametric estimates.

After checking that the published tables are appropriately generated by the replication files, we use the \texttt{MulTE} package \citep{multe2024} to implement the parametric and semiparametric solutions discussed in Section \ref{sec:solutions}. To implement our proposed semiparametric approach, we generate each spline term separately and include it in the regression specification as additional covariates. For brevity, we focus our attention on spline specifications (Equation \eqref{eq:spline-def}) with order $r = 3$ and $J = 2$ knots, where the knots are the 33rd and 67th percentile of each covariate.\footnote{We chose the spline's order and its number of knots arbitrarily and uniformly for all replicated regressions to facilitate inference using the results derived by \cite{Ackerberg2012} and to facilitate comparisons across replicated regressions. Alternatively, we could have used a data-driven procedure to choose these hyperparameters as proposed by \cite{Chen2025}. Since our choice of the spline's order and number of knots does not vary with our replicated studies' sample sizes, one may view our method as flexibly parametric \citep{Spady2025} instead of semiparametric.}

\section{Results}\label{sec:results}

Section \ref{sec:specific-analysis} reports the detailed results regarding the work of \cite{weisburst2019police} while Section \ref{sec:general-analysis} summarizes the results of all replicated regressions. Appendix \ref{ap:tables} reports the detailed results of all replicated articles.

\subsection{In-depth Analysis of Weisburst (2019)} \label{sec:specific-analysis}

To build intuition around the contamination bias present in Equation \eqref{eq:simple-reg} and the parametric and semiparametric solutions discussed in Section \ref{sec:solutions}, we start by discussing the estimates regarding Panel A in Table 2 of \citet{weisburst2019police}. This article analyzes police racial bias in use-of-force incidents in Dallas, Texas, between 2013 and 2016. To replicate this analysis, we implement the following standard regression:
\begin{equation} \label{eq:weisburst}
FR_{a} = \alpha + \beta_{B} \mathbbm{1}(\text{Black})_{a} + \beta_H \mathbbm{1}(\text{Hispanic})_{a} + \beta_O \mathbbm{1}(\text{Other})_{a} + W'_{a} \delta + \lambda_s + \lambda_b + \varepsilon_{a}
\end{equation}
where $a$ indexes an arrestee, the dependent variable indicates whether a use-of-force event occurred during the arrest, and the comparison group consists of White arrestees. Covariates $W_a$ include the arrestee's age and sex, average crime rates in the arrestee's police beat and semester (percentage of violent incidents, property crimes, drug use, prostitution arrests, traffic arrests, and disorderly arrests), and average characteristics of police officers in the arrestee's police beat and semester (age, sex, trainee status, experience, experience squared, Black, Hispanic, or other race). Regression \eqref{eq:weisburst} also includes fixed effects for both beat and semester ($\lambda_b$ and $\lambda_s$). Additionally, we implement the parametric and spline approaches discussed in Section \ref{sec:solutions}.

Table \ref{tab:weisburst} presents our main results for the setting discussed by \cite{weisburst2019police}. Column (1) corresponds to the published estimated coefficients, which are estimated based on Equation \eqref{eq:simple-reg} with the controls and fixed effects described in Equation \eqref{eq:weisburst}. Columns (2), (4) and (6) report the estimates for the linear parametric versions of the ATE solution (Equation \eqref{eq:ate}), the EW solution (Equation \eqref{eq:ew-model}) and the CW solution (Section \ref{SecCW}). Columns (3), (5) and (7) report the estimates for the semiparametric spline versions of the ATE solution (Equation \eqref{eq:ate-spline}), the EW solution (Equation \eqref{eq:ew-spline}) and the CW solution (Section \ref{SecCW}). We report standard errors clustered at the police beat level in parentheses.

\begin{table}
\begin{center}
\caption{Full results for \citet{weisburst2019police}} \label{tab:weisburst}
\setlength{\tabcolsep}{3pt}
\begin{tabular}{@{}l @{}cSSSSS SSSSSS @{}}
    \toprule \hline
    & \multicolumn{7}{c}{Full sample} \\
    \cmidrule(lr){2-8}
    Estimates & \multirow{2}{*}{$\hat{\beta}$} & \multicolumn{2}{c}{ATE} & \multicolumn{2}{c}{EW} & \multicolumn{2}{c}{CW}\\
    \cmidrule(lr){3-4} \cmidrule(lr){5-6} \cmidrule(lr){7-8}
    & & {Linear} & {Spline} & {Linear} & {Spline} & {Linear} & {Spline}\\
    & \text{(1)} & \text{(2)} & \text{(3)} & \text{(4)} & \text{(5)} & \text{(6)} & \text{(7)} \\ \hline
    
Black & 0.172 & 0.342 & 0.540 & 0.109 & 0.061 & 0.246 & 0.660\\
 & (0.274) & (0.396) & (0.952) & (0.267) & (0.289) & (0.292) & (2.281)\\
Hispanic & 0.043 & -0.330 & -0.413 & -0.496 & -0.265 & -0.466 & -0.378\\
 & (0.394) & (0.395) & (0.499) & (0.341) & (0.341) & (0.289) & (1.418)\\
Other & 1.130 & 0.223 & -6.452 & 1.244 & -0.054 & 0.106 & -0.368\\
 & (0.652) & (0.622) & (2.837) & (0.679) & (0.766) & (0.712) & (1.435)\\
 \hline
Number of controls &  & \multicolumn{1}{r}{256} & \multicolumn{1}{r}{310} &  &  &  & \\
Sample size &  & \multicolumn{1}{r}{7,488} & \multicolumn{1}{r}{7,488} &  &  &  & 
\\
    \bottomrule
\end{tabular}
\end{center}
\emph{Notes:} This table reports the coefficients for \citet{weisburst2019police} for all estimators described in Section \ref{sec:solutions}. This table starts by replicating the coefficients from Panel A in Table 2 of the cited article. This article is an observational study. We report standard errors clustered at the police beat level in parentheses.
\end{table}

First, we interpret the original point estimates of Regression \eqref{eq:weisburst}, which are already measured in percentage points. In Column (1), we find that the use-of-force rate is 0.172 higher for Black arrestees, 0.043 higher for Hispanic arrestees, and 1.130 higher for ``other'' arrestees in comparison to White arrestees. 

Second, these estimates may suffer from contamination bias as explained in Section \ref{sec:bias}. The differences between Column (1)'s coefficients and the coefficients in all the other columns provide possible estimates for the contamination bias present in the standard regression. For example, when we compare the estimated coefficient for Black arrestees in the linear specification for the ATE solution against the coefficient in Column (1), we estimate the contamination bias at $-0.170$ (i.e., $= 0.172 - 0.342$), implying that the bias might have a similar magnitude to the estimated coefficient.

We reach similarly concerning point estimates when we compare Column (1)'s coefficients and the coefficients in all the linear parametric specifications (Columns (2), (4), and (6)). However, these differences are small relative to the uncertainty around these estimated coefficients.

When we allow for nonlinear relationships between the potential outcomes and the covariates, the contamination bias problem is even more concerning. For example, for Black arrestees in the ATE solution, we estimate the contamination bias at $-0.368$ (i.e., $= 0.172 - 0.540$). Moreover, the estimated contamination bias is usually larger in magnitude when we compare Column (1)'s coefficients and the coefficients in all the spline semiparametric specifications (Columns (3), (5), and (7)). However, these differences are still small relative to the uncertainty around these estimated coefficients.

Third, note that the spline specification substantially increases the number of control variables. While the linear specification uses 256 control variables, the more flexible spline specification uses 310. This extra flexibility comes at the cost of much larger estimated standard errors in Columns (3), (5), and (7) in comparison to Columns (2), (4), and (6).

The overall message of this replication exercise is that point estimates vary substantially across all specifications. Such large variation might reflect (i) the contamination bias in the standard regression approach, (ii) the extra flexibility that allows spline specification to capture nonlinear relationships, or (iii) the large statistical uncertainty in this empirical setting. To have a better understanding of these issues, we analyze all our replicated regressions in the next section.

\subsection{Analysis of all Selected Papers} \label{sec:general-analysis}

We now present the results for all regressions shown in Table \ref{tab:articles}.\footnote{\label{FtBur}In our estimations, we removed three outliers. The estimates are the ATE estimates of Moral Incentives, Cash Rebate, and Religious Placebo in \citet{bursztyn2019moral}.
The estimates had either abnormally large normalized coefficients or standard error ratios.} Specifically, we analyze a total of 15 standard regressions, 141 parametric estimates, and 141 semiparametric estimates. Given the large number of estimated coefficients, we provide a broad discussion of their point estimates and standard errors.

To allow comparison across papers and against the standard OLS regressions, we implement the normalization procedure proposed by \citet{gphk2024contamination}. We first multiply the estimated coefficient of our six solutions to contamination bias (Section \ref{sec:solutions}) by the sign of the original regression coefficient and then divide it by the standard error of the OLS coefficient. Mathematically, the normalized coefficient is given by $\hat{\beta}_{Norm} = \text{s}(\hat{\beta}_{OLS}) \cdot \hat{\beta}_{TE} / \text{SE} (\hat{\beta}_{OLS})$,
where $\hat{\beta}_{TE}$ is the estimate for one of our six solutions to contamination bias (Section \ref{sec:solutions}) and $\text{s}(\hat{\beta})$ is a function that returns 1 if $\hat{\beta} > 0$, $-1$ if $\hat{\beta} < 0$ and 0 if $\hat{\beta} = 0$.

\begin{figure}[!htb]
\begin{center}
    \begin{subfigure}[t]{0.48\textwidth}
        \includegraphics[width=\textwidth]{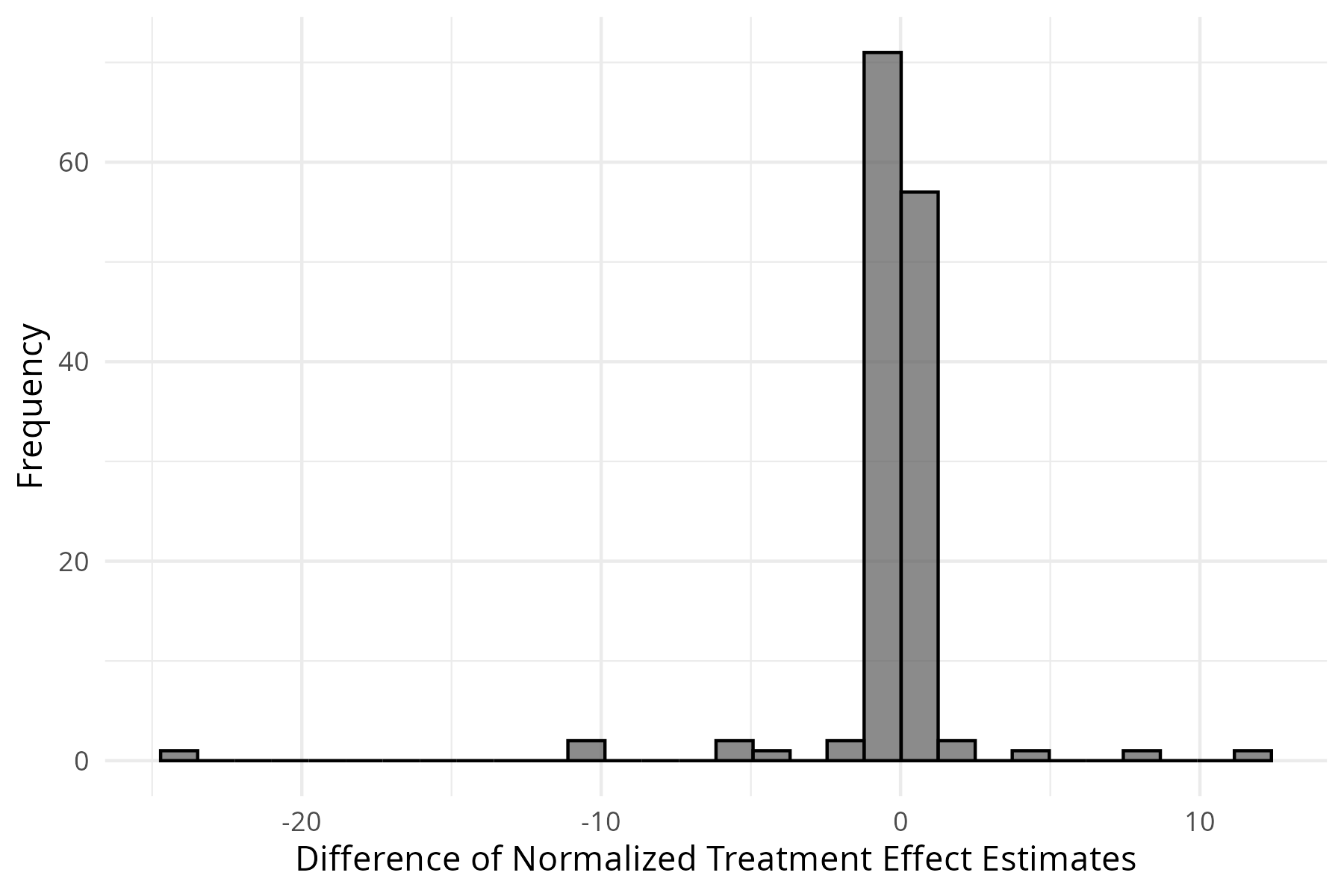}
        \caption{Histogram of the Difference (Spline - Linear) of Normalized Treatment Effects - Full Sample}
        \label{fig:histogram-full}
    \end{subfigure}
\hfill
    \begin{subfigure}[t]{0.48\textwidth}
        \includegraphics[width=\textwidth]{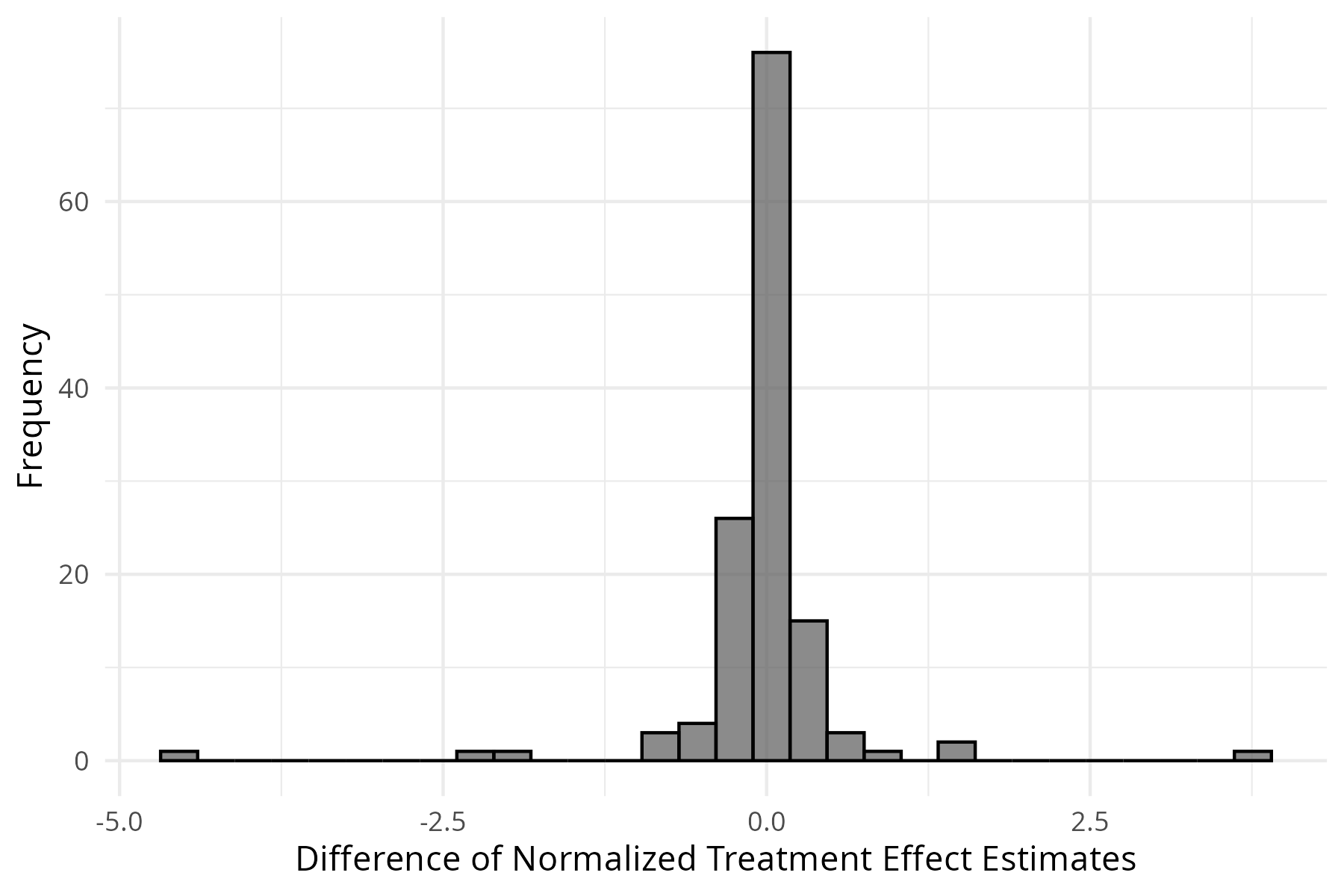}
        \caption{Histogram of the Difference (Spline - Linear) of Normalized Treatment Effects - Zoomed In Version}
        \label{fig:histogram}
    \end{subfigure}
\end{center}
\caption{Histogram of the Difference (Spline - Linear) of Normalized Treatment Effects}\label{FigHist}
\textit{Notes:} This figure presents histograms of the difference of normalized coefficients, $\hat{\beta}_{NormDif}$, for each treatment effect estimator in Section \ref{sec:solutions}. The normalized coefficient is given by $\hat{\beta}_{Norm} = \text{s}(\hat{\beta}_{OLS}) \cdot \hat{\beta}_{TE} / \text{SE} (\hat{\beta}_{OLS})$, where $\text{s}(\hat{\beta})$ is the sign function and $\hat{\beta}_{TE}$ is one of the estimators. The difference between the normalized coefficients is defined as $\hat{\beta}_{NormDif} = \hat{\beta}^{Spline}_{Norm} - \hat{\beta}^{Linear}_{Norm}$. Figure \ref{fig:histogram} focuses on normalized coefficients between $-5$ and 5 while Figure \ref{fig:histogram-full} uses the full sample after excluding the outliers mentioned in Footnote \ref{FtBur}.
\end{figure}

After normalizing all estimated coefficients, we take the difference between the spline-based ones and the linear ones. Specifically, for each effect of interest, we compute $\hat{\beta}_{Norm Dif} = \hat{\beta}^{Spline}_{Norm} - \hat{\beta}^{Linear}_{Norm}$. Due to the previous normalization procedure, $\hat{\beta}_{Norm Dif}$ is comparable across all regressions and provides an easy-to-interpret measure of the distance between the two point estimates. Figure \ref{fig:histogram-full} contains a histogram of $\hat{\beta}_{Norm Dif}$ while Figure \ref{fig:histogram} shows a zoomed-in version of this histogram, which focuses on normalized differences between $-5$ and 5.

Initially, note that linear parametric estimates and semiparametric spline estimates are not always similar to each other. If they were, the histograms in Figure \ref{FigHist} would be a spike at 0, since we divide both types of estimates by the same denominator in the normalization process. This result suggests that allowing for nonlinear relationships between the potential outcomes and the covariates may impact the estimated treatment effect parameter in real empirical settings.

Next, we must ask whether these differences between linear parametric estimates and semiparametric spline estimates are substantial. To answer this question, we can benchmark our normalized differences against two possible thresholds. First, we measure these differences in standard errors of the original OLS estimate. For example, a difference greater than 0.2 might be regarded as large since it implies that the difference between the parametrically and semiparametrically estimated coefficients is the size of one-fifth of the original OLS standard error. Importantly, almost 30\% (42/141) of our estimated differences are above this threshold. Second, we can benchmark these differences against the standard normal distribution. For example, we find that 11 estimated differences exceed the 95th percentile of the standard normal, i.e., 7.8\% of our estimated differences are larger in absolute value than 1.645. These two benchmarks indicate substantial differences between the linear parametric estimators proposed by \cite{gphk2024contamination} and the semiparametric spline estimators suggested in Section \ref{sec:solutions}.

Lastly, we must ask whether allowing nonlinear relationships in our spline specifications comes at the cost of increased statistical uncertainty. To answer this question, we analyze two different ratios: one between the estimated standard error of all our six solutions to contamination bias (Section \ref{sec:solutions}) and the estimated standard error of the original OLS regression (Equation \eqref{eq:simple-reg}), and the other between the semiparametric spline estimates and the parametric linear estimates for each of the six solutions. For the first ratio, mathematically, we calculate $\sfrac{SE_{TE}}{SE_{OLS}}$. Table \ref{tab:quantiles} shows the quantiles of these six ratios. Each pair of rows is associated with one of the three solutions (ATE, EW, or CW) and, within each pair, the first row is associated with semiparametric spline estimators while the second row is associated with linear parametric estimators.

\begin{table}[!htbp]
\centering
\caption{\label{tab:quantiles} Quantiles of the SE Ratio for Each Parametric and Semiparametric Solution}
\begin{tabular}[t]{ccccccccc}
\toprule
\multirow{2}{*}{Estimator} & \multirow{2}{*}{Spline} & Min & 10\% & 25\% & 50\% & 75\% & 90\% & Max\\
& & (1) & (2) & (3) & (4) & (5) & (6) & (7) \\
\midrule
 & \checkmark & 0.595 & 0.818 & 0.958 & 0.984 & 1.986 & 15.716 & 60.852\\
\cmidrule{2-9}
\multirow{-2}{*}{ATE} & \xmark & 0.596 & 0.815 & 0.927 & 1.001 & 1.011 & 1.312 & 10.374\\
\cmidrule{1-9}
 & \checkmark & 0.730 & 0.902 & 0.972 & 0.998 & 1.004 & 1.064 & 1.242\\
\cmidrule{2-9}
\multirow{-2}{*}{EW} & \xmark & 0.646 & 0.899 & 0.959 & 1.000 & 1.003 & 1.031 & 1.086\\
\cmidrule{1-9}
 & \checkmark & 0.837 & 0.917 & 0.982 & 1.015 & 1.055 & 1.549 & 8.430\\
\cmidrule{2-9}
\multirow{-2}{*}{CW} & \xmark & 0.732 & 0.934 & 0.985 & 1.008 & 1.042 & 1.235 & 2.079\\
\bottomrule
\end{tabular} \\
\RaggedRight
\textit{Notes:}  Table \ref{tab:quantiles} shows the quantiles (specifically, the 0, 10, 25, 50, 75, 90 and 100 percentiles) of ratios between the estimated standard error of all our six solutions to contamination bias (Section \ref{sec:solutions}) and the estimated standard error of the original OLS regression (Equation \eqref{eq:simple-reg}). Each pair of rows is associated with one of the three solutions (ATE, EW, or CW) and, within each pair, the first row is associated with semiparametric spline estimators while the second row is associated with linear parametric estimators.
\end{table}

First, we compare the same solution using linear parametric estimators versus semiparametric spline estimators. Unsurprisingly, almost all of the percentiles in the spline rows are larger than the percentiles in the linear rows.

Second, and most interestingly, most estimated standard errors for the estimators that address contamination bias are not much larger than the estimated standard error of the usual regression in Equation \eqref{eq:simple-reg}. For example, the 50th percentile is around 1 for all specifications, implying that the estimated standard errors of half of the coefficients for all six solutions in Section \ref{sec:solutions} are smaller than the estimated standard errors for the usual regression in Equation \eqref{eq:simple-reg}. Additionally, the highest standard error of the EW solution is only 24\% bigger than the standard error of its comparable OLS estimator. These results imply that (i) adopting approaches that address contamination bias concerns and (ii) allowing for nonlinear relationships between potential outcomes and covariates may not lead to increased statistical uncertainty, especially for the EW and CW approaches.\footnote{The last three columns of Table \ref{tab:quantiles} suggest that the ATE approach produces large increases in uncertainty in some scenarios.}

We now turn to the second ratio of estimated standard errors. Mathematically, we calculate $\sfrac{SE_{Spl}^{TE}}{SE_{Lin}^{TE}}$. The results are in Table \ref{tab:TE-quantiles}, and they support the claim that allowing nonlinear relationships does not increase uncertainty substantially. As in Table \ref{tab:quantiles}, the median of each ratio distribution is around 1 for all three estimators, implying that half of the semiparametric spline estimates yield smaller standard errors than the parametric linear standard errors. This result implies that, even between the contamination-bias solutions, the semiparametric estimates do not incur substantially greater uncertainty costs, in particular for the EW and CW approaches.\footnote{The same may not be said for the ATE solution, as the last three columns of Table \ref{tab:quantiles} suggest it may produce large increases in uncertainty in some scenarios.}.

\begin{table}[!htbp]
\centering
\caption{\label{tab:TE-quantiles} Quantiles of the Spline SE Ratio for each Solution}
\begin{tabular}[t]{cccccccc}
\toprule
\multirow{2}{*}{Estimator} & Min & 10\% & 25\% & 50\% & 75\% & 90\% & Max\\
& (1) & (2) & (3) & (4) & (5) & (6) & (7) \\
\midrule
ATE & 0.723 & 0.930 & 0.964 & 1.001 & 1.261 & 16.198 & 59.518\\
\cmidrule{1-8}
EW & 0.929 & 0.966 & 0.986 & 0.998 & 1.001 & 1.032 & 1.134\\
\cmidrule{1-8}
CW & 0.892 & 0.939 & 0.978 & 1.000 & 1.012 & 1.032 & 7.809\\
\bottomrule
\end{tabular} \\
\RaggedRight
\textit{Notes:}  Table \ref{tab:TE-quantiles} shows the quantiles (specifically, the 0, 10, 25, 50, 75, 90 and 100 percentiles) of ratios between the semiparametric spline and parametric linear estimated standard errors for each contamination bias solution (Section \ref{sec:solutions}). 
\end{table}

\section{Conclusion}\label{sec:conclusion}

Multi-valued discrete treatments are widespread in observational studies and randomized controlled trials. \cite{gphk2024contamination} show that standard regression approaches do not estimate a proper weighted average of treatment effects, presenting contamination bias. They discuss three alternative estimators that avoid this bias and implement their parametric versions.

In this paper, we deepen their analysis by explaining and implementing three semiparametric versions of their solutions to contamination bias. We find that (i) contamination bias might be a relevant problem in many empirical settings, (ii) there exist substantial differences between the point estimates of linear parametric estimators and semiparametric spline estimators, and (iii) adopting flexible semiparametric solutions to contamination bias may not lead to increased statistical uncertainty. Because our proposed spline estimators are easy to implement using the statistical package provided by \cite{multe2024}, we recommend that empirical researchers verify the robustness of their conclusions to the use of semiparametric tools that address contamination bias.

As future avenues of research, it is important to explore how machine learning estimators \citep{Cetin2025} behave in empirical settings with multi-valued discrete treatments.


\singlespace

\bibliography{references_part2}


\pagebreak

\newpage

\pagebreak

\setcounter{table}{0}
\renewcommand\thetable{A.\arabic{table}}

\setcounter{figure}{0}
\renewcommand\thefigure{A.\arabic{figure}}

\setcounter{equation}{0}
\renewcommand\theequation{A.\arabic{equation}}

\appendix

\begin{center}
	\huge
	Supporting Information

	(Online Appendix)

\end{center}

\doublespacing
\normalsize


\setcounter{table}{0}
\renewcommand\thetable{A.\arabic{table}}

\setcounter{figure}{0}
\renewcommand\thefigure{A.\arabic{figure}}

\setcounter{equation}{0}
\renewcommand\theequation{A.\arabic{equation}}

\setcounter{theorem}{0}
\renewcommand\thetheorem{A.\arabic{theorem}}

\setcounter{proposition}{0}
\renewcommand\theproposition{A.\arabic{proposition}}

\setcounter{corollary}{0}
\renewcommand\thecorollary{A.\arabic{corollary}}

\setcounter{assumption}{0}
\renewcommand\theassumption{A.\arabic{assumption}}

\setcounter{definition}{0}
\renewcommand\thedefinition{A.\arabic{definition}}

\setcounter{Lemma}{0}
\renewcommand\theLemma{A.\arabic{Lemma}}

\section{Additional Tables} \label{ap:tables}

\subsection{Articles Selected for Replication}

When selecting articles for replication, we searched for articles published by the American Economic Association and in Econometrica, the Journal of Political Economy, the Quarterly Journal of Economics, and the Review of Economic Studies. Each journals' website was manually searched in the period of 2010-2024, using the methodology described in detail in Section \ref{sec:method}. The initial search round yielded 19 papers, in addition to the 9 papers replicated in \citet{gphk2024contamination}, for a total of 26 replicable papers. Although we found articles published in the Review of Economic Studies and the Quarterly Journal of Economics, we did not include them (and some articles from the other journals) in our sample of replicated articles for four reasons:
\begin{enumerate}
    \item their data were only partially available (4 articles);
    \item they did not in fact contain multi-valued discrete treatments (6 articles);
    \item they did not contain any continuous covariates to justify the use of spline estimators (4 articles); or
    \item they did not satisfy the overlap restriction when we implemented our spline estimators (1 article).
\end{enumerate}

Our final sample contains 11 articles, from the initial 26. The list, and which tables were replicated, can be found in Table \ref{tab:articles}, below.

\begin{table}[!htb]
    \caption{Articles Selected for Replication}\label{tab:articles}
\begin{center}
\begin{tabular}{l c c c }
\hline \hline
\multirow{2}{*}{Article} & Tables & \multirow{2}{*}{RCT} & Analyzed \\
 & (Columns) & & by GHK (2024) \\
\hline
 \citet{jha2019valuing} & 9 (1 - Panels A, B) & $\checkmark$ & \\
 \citet{bursztyn2019moral} & 3-6 (3) & $\checkmark$ & \\ 
 \citet{brownback2020instruction} & 2 (2) & $\checkmark$ &  \\ 
 \citet{lee2020electrification} & 2 (2) & $\checkmark$ &  \\
 \citet{dellavigne2022preferences} & 1 (2, 3, 5, 6) & $\checkmark$ & \\ \hline
 \citet{drexler2014financial} &  2 (row 2) & $\checkmark$ & $\checkmark$ \\
 \citet{duflo2015education} & 2 (1, Panel A) & $\checkmark$ & $\checkmark$ \\
 \citet{benhassine2015shove} &  5 (1) & $\checkmark$ & $\checkmark$ \\ 
 \citet{fryerlevitt2013racial} & 3 (4) &  & $\checkmark$ \\
 \citet{weisburst2019police} &  2 (Panel A) &  & $\checkmark$ \\ 
 \citet{rim2020disparities} &  2 (3) &  & $\checkmark$ \\ 
\hline
\end{tabular}
\end{center}
\textit{Notes}: This table summarizes the replicated articles, specifying which of their tables and columns were analyzed. It also indicates which articles are randomized control trials (RCT) or observational studies and which articles were previously analyzed by \cite{gphk2024contamination}.
\end{table}

\newpage

\subsection{Detailed Results for all Replicated Articles}
    
In this Appendix, we present all replication results tables.
The regressions analyzed are specified in the last column of Table \ref{tab:articles}.
Table 2, Panel A from \citet{weisburst2019police}  is presented in the main text, and not in this appendix.

In Tables \ref{tab:art-3-t2-c2}-\ref{tab:rim}, we report the estimated coefficients from each estimator presented in Section \ref{sec:solutions}.
The Tables report the estimates for the simple OLS and each solution's estimator, with and without splines.
If overlap fails in the analysis, we follow the \texttt{MulTE} package: we drop covariates with no treatment variation.
By subsetting, it is possible to estimate the ATE.
In the tables that overlap failed, we report the OLS, EW, and CW estimates for the full sample (with and without splines), and all 7 estimators for the subsetted sample.
Furthermore, when there is more than one replication for an article, we specify which specific regression the table refers to in the caption and the notes.

\newpage

\begin{table}
\centering
\begin{threeparttable}
\caption{Full results: \citet{dellavigne2022preferences}, Table 1 - Column 2} \label{tab:art-3-t2-c2}
\setlength{\tabcolsep}{3pt}
\begin{tabular}{@{}l @{}cSSSSS SSSSSS @{}}
    \toprule \hline
    & \multicolumn{7}{c}{Full sample} \\
    \cmidrule(lr){2-8}
    Estimates & \multirow{2}{*}{$\hat{\beta}$} & \multicolumn{2}{c}{ATE} & \multicolumn{2}{c}{EW} & \multicolumn{2}{c}{CW}\\
    \cmidrule(lr){3-4} \cmidrule(lr){5-6} \cmidrule(lr){7-8}
    & & {Linear} & {Spline} & {Linear} & {Spline} & {Linear} & {Spline}\\ \hline
    
Positive Gift & 0.903 & 0.906 & 0.841 & 0.905 & 0.833 & 0.931 & 0.708\\
 & (0.736) & (0.735) & (0.709) & (0.734) & (0.726) & (0.733) & (0.748)\\
Negative Gift & -0.014 & -0.015 & -0.079 & -0.014 & -0.068 & -0.035 & -0.135\\
 & (0.744) & (0.742) & (0.702) & (0.744) & (0.723) & (0.748) & (0.721)\\
Positive In-Kind Gift & -1.011 & -1.012 & -1.014 & -1.013 & -1.074 & -1.030 & -1.125\\
 & (0.971) & (0.969) & (0.943) & (0.974) & (0.964) & (0.981) & (0.968)\\
\hline
Number of controls &  & \multicolumn{1}{r}{ 1} & \multicolumn{1}{r}{ 5} &  &  &  & \\
Sample size &  & \multicolumn{1}{r}{892} & \multicolumn{1}{r}{892} &  &  &  & 
\\
    \bottomrule
\end{tabular}
\begin{tablenotes}\item  \emph{Notes:} This table reports the coefficients for \citet{dellavigne2022preferences} for all estimators described in Section \ref{sec:solutions}. This table starts by replicating the coefficients from Column 2 in Table 1 of the cited article. This is an experimental paper. We report standard errors clustered at the session level in parentheses.
\end{tablenotes}
\end{threeparttable}
\end{table}
  
\begin{table}
\centering
\begin{threeparttable}
\caption{Full results: \citet{dellavigne2022preferences}, Table 1 - Column 3} \label{tab:art-3-t2-c3}

\setlength{\tabcolsep}{3pt}
\begin{tabular}{@{}l @{}cSSSSS SSSSSS @{}}
    \toprule \hline
    & \multicolumn{7}{c}{Full sample} \\
    \cmidrule(lr){2-8}
    Estimates & \multirow{2}{*}{$\hat{\beta}$} & \multicolumn{2}{c}{ATE} & \multicolumn{2}{c}{EW} & \multicolumn{2}{c}{CW}\\
    \cmidrule(lr){3-4} \cmidrule(lr){5-6} \cmidrule(lr){7-8}
    & & {Linear} & {Spline} & {Linear} & {Spline} & {Linear} & {Spline}\\ \hline
    
Positive Gift & 0.603 & 0.601 & 0.476 & 0.601 & 0.526 & 0.613 & 0.220\\
 & (0.727) & (0.728) & (0.705) & (0.727) & (0.718) & (0.725) & (0.756)\\
Negative Gift & -0.047 & -0.054 & -0.130 & -0.047 & -0.075 & -0.053 & -0.270\\
 & (0.752) & (0.755) & (0.708) & (0.753) & (0.724) & (0.755) & (0.727)\\
Positive In-Kind Gift & -1.090 & -1.092 & -0.517 & -1.093 & -1.218 & -1.097 & -1.511\\
 & (0.925) & (0.930) & (0.918) & (0.930) & (0.900) & (0.931) & (0.947)\\
\hline
Number of controls &  & \multicolumn{1}{r}{ 1} & \multicolumn{1}{r}{ 5} &  &  &  & \\
Sample size &  & \multicolumn{1}{r}{892} & \multicolumn{1}{r}{892} &  &  &  & 
\\
    \bottomrule
\end{tabular}
\begin{tablenotes}\item  \emph{Notes:} This table reports the coefficients for \citet{dellavigne2022preferences} for all estimators described in Section \ref{sec:solutions}. This table starts by replicating the coefficients from Column 3 in Table 1 of the cited article. This is an experimental paper. We report standard errors clustered at the session level in parentheses.
  \end{tablenotes}
 \end{threeparttable}
\end{table}

\newpage

\begin{table}
\centering
\begin{threeparttable} 
\caption{Full results: \citet{dellavigne2022preferences}, Table 1 - Column 5} \label{tab:art-3-t2-c5}
\setlength{\tabcolsep}{3pt}
\begin{tabular}{@{}l @{}cSSSSS SSSSSS @{}}
    \toprule \hline
    & \multicolumn{7}{c}{Full sample} \\
    \cmidrule(lr){2-8}
    Estimates & \multirow{2}{*}{$\hat{\beta}$} & \multicolumn{2}{c}{ATE} & \multicolumn{2}{c}{EW} & \multicolumn{2}{c}{CW}\\
    \cmidrule(lr){3-4} \cmidrule(lr){5-6} \cmidrule(lr){7-8}
    & & {Linear} & {Spline} & {Linear} & {Spline} & {Linear} & {Spline}\\ \hline
    
Positive Gift & 0.024 & 0.024 & 0.021 & 0.024 & 0.021 & 0.025 & 0.021\\
 & (0.022) & (0.022) & (0.021) & (0.022) & (0.022) & (0.022) & (0.022)\\
Negative Gift & -0.014 & -0.014 & -0.013 & -0.014 & -0.015 & -0.015 & -0.013\\
 & (0.031) & (0.031) & (0.028) & (0.031) & (0.032) & (0.031) & (0.030)\\
Positive In-Kind Gift & -0.033 & -0.033 & -0.036 & -0.033 & -0.035 & -0.034 & -0.035\\
 & (0.029) & (0.029) & (0.028) & (0.029) & (0.029) & (0.029) & (0.029)\\
\hline
Number of controls &  & \multicolumn{1}{r}{ 1} & \multicolumn{1}{r}{ 5} &  &  &  & \\
Sample size &  & \multicolumn{1}{r}{892} & \multicolumn{1}{r}{892} &  &  &  & 
\\
    \bottomrule
\end{tabular}
\begin{tablenotes}\item  \emph{Notes:}  This table reports the coefficients for \citet{dellavigne2022preferences} for all estimators described in Section \ref{sec:solutions}. This table starts by replicating the coefficients from Column 5 in Table 1 of the cited article. This is an experimental paper. We report standard errors clustered at the session level in parentheses.
  \end{tablenotes}
 \end{threeparttable}
\end{table}

\begin{table}
\centering
\begin{threeparttable} 
\caption{Full results: \citet{dellavigne2022preferences}, Table 1 - Column 6} \label{tab:art-3-t2-c6}
\setlength{\tabcolsep}{3pt}
\begin{tabular}{@{}l @{}cSSSSS SSSSSS @{}}
    \toprule \hline
    & \multicolumn{7}{c}{Full sample} \\
    \cmidrule(lr){2-8}
    Estimates & \multirow{2}{*}{$\hat{\beta}$} & \multicolumn{2}{c}{ATE} & \multicolumn{2}{c}{EW} & \multicolumn{2}{c}{CW}\\
    \cmidrule(lr){3-4} \cmidrule(lr){5-6} \cmidrule(lr){7-8}
    & & {Linear} & {Spline} & {Linear} & {Spline} & {Linear} & {Spline}\\ \hline
    
Positive Gift & 0.016 & 0.016 & 0.013 & 0.016 & 0.013 & 0.017 & 0.014\\
 & (0.021) & (0.021) & (0.020) & (0.021) & (0.021) & (0.021) & (0.021)\\
Negative Gift & -0.018 & -0.018 & -0.017 & -0.018 & -0.015 & -0.018 & -0.016\\
 & (0.031) & (0.031) & (0.029) & (0.031) & (0.029) & (0.031) & (0.029)\\
Positive In-Kind Gift & -0.035 & -0.035 & -0.035 & -0.035 & -0.035 & -0.035 & -0.035\\
 & (0.028) & (0.028) & (0.027) & (0.028) & (0.027) & (0.028) & (0.028)\\
\hline
Number of controls &  & \multicolumn{1}{r}{ 1} & \multicolumn{1}{r}{ 5} &  &  &  & \\
Sample size &  & \multicolumn{1}{r}{892} & \multicolumn{1}{r}{892} &  &  &  & 
\\
    \bottomrule
\end{tabular}
\begin{tablenotes}\item  \emph{Notes:}  This table reports the coefficients for \citet{dellavigne2022preferences} for all estimators described in Section \ref{sec:solutions}. This table starts by replicating the coefficients from Column 6 in Table 1 of the cited article. This is an experimental paper. We report standard errors clustered at the session level in parentheses.
  \end{tablenotes}
 \end{threeparttable}
\end{table}

\newpage

\begin{landscape}
\begin{table}
\begin{threeparttable}
\caption{Full results: \cite{jha2019valuing}, Table 9 - Panel A - Column 1} \label{tab:art-8-PA-c1}
\setlength{\tabcolsep}{3pt}
\begin{tabular}{@{}l @{}cSSSS cSSSSSS @{}}
    \toprule \hline
    & \multicolumn{5}{c}{Full sample} & \multicolumn{7}{c}{Overlap} \\
    \cmidrule(lr){2-6} \cmidrule(lr){7-13}
    Estimates & \multirow{2}{*}{$\hat{\beta}$} & \multicolumn{2}{c}{EW} & \multicolumn{2}{c}{CW} &
                \multirow{2}{*}{$\hat{\beta}$} & \multicolumn{2}{c}{ATE} & \multicolumn{2}{c}{EW} & \multicolumn{2}{c}{CW} \\
    \cmidrule(lr){3-4} \cmidrule(lr){5-6} \cmidrule(lr){8-9} \cmidrule(lr){10-11} \cmidrule(lr){12-13}
    & & {Linear} & {Spline} & {Linear} & {Spline} &
      & {Linear} & {Spline} & {Linear} & {Spline} & {Linear} & {Spline}\\ \hline
    
Palestinian Assets & 0.032 &  0.025 & 0.022 & 0.028 & 0.024 & 0.034 & 0.030 & 0.019 & 0.028 & 0.024 & 0.031 & 0.026\\
 & (0.021) &  (0.020) & (0.020) & (0.020) & (0.020) & (0.021) & (0.019) & (0.020) & (0.020) & (0.020) & (0.020) & (0.020)\\
Non-Palestinian Assets & 0.065 &  0.067 & 0.064 & 0.070 & 0.065 & 0.067 & 0.069 & 0.054 & 0.069 & 0.067 & 0.072 & 0.067\\
 & (0.019) &  (0.019) & (0.018) & (0.019) & (0.019) & (0.019) & (0.018) & (0.018) & (0.019) & (0.019) & (0.019) & (0.019)\\
\cmidrule(lr){2-6} \cmidrule(lr){7-13}
Number of controls &  & \multicolumn{1}{r}{131} & \multicolumn{1}{r}{146} &  &  &  & \multicolumn{1}{r}{129} & \multicolumn{1}{r}{144} &  &  &  & \\
Sample size &  & \multicolumn{1}{r}{1,311} & \multicolumn{1}{r}{1,311} &  &  &  & \multicolumn{1}{r}{1,305} & \multicolumn{1}{r}{1,305} &  &  &  & 
\\
    \bottomrule
\end{tabular}
\begin{tablenotes}\item\emph{Notes:} This table reports the coefficients for \citet{jha2019valuing} for all estimators described in Section \ref{sec:solutions}. This table starts by replicating the coefficients from Panel A, Column 1 in Table 9 of the cited article. Since overlap failed, we report both sets of estimands: full and overlap sample. This article is an experimental study. We report heteroskedasticity-robust standard errors in parentheses.
\end{tablenotes}
 \end{threeparttable}
\end{table}

\begin{table}
\begin{threeparttable}
\caption{Full results: \citet{jha2019valuing}, Table 9 - Panel B - Column 1} \label{tab:art-8-PB-c1}
\setlength{\tabcolsep}{3pt}
\begin{tabular}{@{}l @{}cSSSS cSSSSSS @{}}
    \toprule \hline
    & \multicolumn{5}{c}{Full sample} & \multicolumn{7}{c}{Overlap} \\
    \cmidrule(lr){2-6} \cmidrule(lr){7-13}
    Estimates & \multirow{2}{*}{$\hat{\beta}$} & \multicolumn{2}{c}{EW} & \multicolumn{2}{c}{CW} &
                \multirow{2}{*}{$\hat{\beta}$} & \multicolumn{2}{c}{ATE} & \multicolumn{2}{c}{EW} & \multicolumn{2}{c}{CW} \\
    \cmidrule(lr){3-4} \cmidrule(lr){5-6} \cmidrule(lr){8-9} \cmidrule(lr){10-11} \cmidrule(lr){12-13}
    & & {Linear} & {Spline} & {Linear} & {Spline} &
      & {Linear} & {Spline} & {Linear} & {Spline} & {Linear} & {Spline}\\ \hline
    
Palestinian Assets & 0.111 &  0.103 & 0.127 & 0.097 & 0.119 & 0.116 & 0.109 & 0.150 & 0.108 & 0.133 & 0.102 & 0.124\\
 & (0.049) &  (0.046) & (0.046) & (0.045) & (0.046) & (0.049) & (0.044) & (0.046) & (0.046) & (0.046) & (0.045) & (0.046)\\
Non-Palestinian Assets & 0.110 &  0.112 & 0.109 & 0.105 & 0.105 & 0.111 & 0.107 & 0.115 & 0.114 & 0.111 & 0.108 & 0.109\\
 & (0.045) &  (0.044) & (0.044) & (0.042) & (0.043) & (0.045) & (0.041) & (0.042) & (0.044) & (0.044) & (0.042) & (0.043)\\
\cmidrule(lr){2-6} \cmidrule(lr){7-13}
Number of controls &  & \multicolumn{1}{r}{131} & \multicolumn{1}{r}{146} &  &  &  & \multicolumn{1}{r}{129} & \multicolumn{1}{r}{144} &  &  &  & \\
Sample size &  & \multicolumn{1}{r}{1,277} & \multicolumn{1}{r}{1,277} &  &  &  & \multicolumn{1}{r}{1,271} & \multicolumn{1}{r}{1,271} &  &  &  & 
\\
    \bottomrule
\end{tabular}
\begin{tablenotes}\item\emph{Notes:} This table reports the coefficients for \citet{jha2019valuing} for all estimators described in Section \ref{sec:solutions}. This table starts by replicating the coefficients from Panel B, Column 1 in Table 9 of the cited article. Since overlap failed, we report both sets of estimands: full and overlap sample. This article is an experimental study. We report heteroskedasticity-robust standard errors in parentheses.
\end{tablenotes}
 \end{threeparttable}
\end{table}
\end{landscape}

\newpage

\begin{landscape}
\begin{table}
\begin{threeparttable}
\caption{Full results: \citet{bursztyn2019moral}, Tables 2-6 - Column 3} \label{tab:art-10}
\setlength{\tabcolsep}{3pt}
\begin{tabular}{@{}l @{}cSSSS cSSSSSS @{}}
    \toprule \hline
    & \multicolumn{5}{c}{Full sample} & \multicolumn{7}{c}{Overlap} \\
    \cmidrule(lr){2-6} \cmidrule(lr){7-13}
    Estimates & \multirow{2}{*}{$\hat{\beta}$} & \multicolumn{2}{c}{EW} & \multicolumn{2}{c}{CW} &
                \multirow{2}{*}{$\hat{\beta}$} & \multicolumn{2}{c}{ATE} & \multicolumn{2}{c}{EW} & \multicolumn{2}{c}{CW} \\
    \cmidrule(lr){3-4} \cmidrule(lr){5-6} \cmidrule(lr){8-9} \cmidrule(lr){10-11} \cmidrule(lr){12-13}
    & & {Linear} & {Spline} & {Linear} & {Spline} &
      & {Linear} & {Spline} & {Linear} & {Spline} & {Linear} & {Spline}\\ \hline
    
Moral Incentive & -0.051 &  -0.052 & -0.052 &  &  & -0.039 & -0.037 & -0.532 & -0.039 & -0.040 & -0.039 & -0.038\\
 & (0.013) &  (0.013) & (0.013) &  &  & (0.012) & (0.012) & (3.123) & (0.012) & (0.012) & (0.013) & (0.013)\\
Cash Rebate & -0.003 &  -0.015 & -0.010 &  &  & 0.028 & 0.019 & -6.812 & 0.026 & 0.028 & 0.018 & 0.019\\
 & (0.029) &  (0.035) & (0.035) &  &  & (0.026) & (0.027) & (8.946) & (0.026) & (0.026) & (0.027) & (0.027)\\
Credit Reputation & -0.104 &  -0.103 & -0.104 &  &  & -0.118 & -0.121 & -0.050 & -0.117 & -0.117 & -0.123 & -0.123\\
 & (0.013) &  (0.014) & (0.014) &  &  & (0.013) & (0.013) & (0.794) & (0.013) & (0.013) & (0.013) & (0.013)\\
Simple Reminder & -0.022 &  -0.019 & -0.017 &  &  & -0.019 & -0.016 & 0.064 & -0.019 & -0.016 & -0.018 & -0.012\\
 & (0.015) &  (0.015) & (0.015) &  &  & (0.015) & (0.015) & (0.794) & (0.015) & (0.015) & (0.015) & (0.015)\\
Religious Placebo & -0.010 &  -0.007 & -0.007 &  &  & -0.023 & -0.018 & 7.509 & -0.019 & -0.018 & -0.022 & -0.019\\
 & (0.017) &  (0.017) & (0.017) &  &  & (0.016) & (0.017) & (9.470) & (0.016) & (0.016) & (0.017) & (0.017)\\
Implicit Moral Incentive & -0.039 &  -0.038 & -0.037 &  &  & -0.007 & -0.013 & 0.059 & -0.009 & -0.010 & -0.012 & -0.011\\
 & (0.018) &  (0.019) & (0.019) &  &  & (0.016) & (0.016) & (0.794) & (0.016) & (0.016) & (0.016) & (0.016)\\
Nonreligious Moral Incentive & -0.038 &  -0.038 & -0.040 &  &  & -0.007 & 0.003 & 0.160 & -0.007 & -0.010 & -0.002 & -0.003\\
 & (0.017) &  (0.019) & (0.019) &  &  & (0.015) & (0.015) & (0.809) & (0.015) & (0.015) & (0.015) & (0.015)\\
\cmidrule(lr){2-6} \cmidrule(lr){7-13}
Number of controls &  & \multicolumn{1}{r}{45} & \multicolumn{1}{r}{56} &  &  &  & \multicolumn{1}{r}{34} & \multicolumn{1}{r}{45} &  &  &  & \\
Sample size &  & \multicolumn{1}{r}{13,428} & \multicolumn{1}{r}{13,428} &  &  &  & \multicolumn{1}{r}{13,334} & \multicolumn{1}{r}{13,334} &  &  &  & 
\\
    \bottomrule
\end{tabular}
\begin{tablenotes}\item\emph{Notes:} This table reports the coefficients for \citet{bursztyn2019moral} for all estimators described in Section \ref{sec:solutions}. This table starts by replicating the coefficients from Column 3 in Tables 2-6 of the cited article. Since overlap failed, we report both sets of estimands: full and overlap sample. Overlap failure also explains why the $CW$ Columns are empty. This article is an experimental study. We report heteroskedasticity-robust standard errors in parentheses.
\end{tablenotes}
\end{threeparttable}
\end{table}
\end{landscape}

\newpage

\begin{landscape}
\begin{table}
\begin{threeparttable}
\caption{Full results: \citet{brownback2020instruction}, Table 2 - Column 2} \label{tab:art-13-t2-c2}
\setlength{\tabcolsep}{3pt}
\begin{tabular}{@{}l @{}cSSSS cSSSSSS @{}}
    \toprule \hline
    & \multicolumn{5}{c}{Full sample} & \multicolumn{7}{c}{Overlap} \\
    \cmidrule(lr){2-6} \cmidrule(lr){7-13}
    Estimates & \multirow{2}{*}{$\hat{\beta}$} & \multicolumn{2}{c}{EW} & \multicolumn{2}{c}{CW} &
                \multirow{2}{*}{$\hat{\beta}$} & \multicolumn{2}{c}{ATE} & \multicolumn{2}{c}{EW} & \multicolumn{2}{c}{CW} \\
    \cmidrule(lr){3-4} \cmidrule(lr){5-6} \cmidrule(lr){8-9} \cmidrule(lr){10-11} \cmidrule(lr){12-13}
    & & {Linear} & {Spline} & {Linear} & {Spline} &
      & {Linear} & {Spline} & {Linear} & {Spline} & {Linear} & {Spline}\\ \hline
    
Instructor Incentives & 0.161 &  0.148 & 0.152 & 0.288 & 0.280 & 0.172 & 0.182 & 0.191 & 0.175 & 0.181 & 0.208 & 0.179\\
 & (0.060) &  (0.058) & (0.058) & (0.105) & (0.106) & (0.063) & (0.069) & (0.071) & (0.062) & (0.062) & (0.067) & (0.069)\\
Combined Incentives & 0.099 &  0.107 & 0.109 & 0.292 & 0.278 & 0.183 & 0.277 & 0.260 & 0.210 & 0.196 & 0.247 & 0.177\\
 & (0.083) &  (0.094) & (0.094) & (0.095) & (0.093) & (0.077) & (0.080) & (0.076) & (0.080) & (0.080) & (0.078) & (0.070)\\
Student Incentives & -0.052 & 0.012 & 0.009 & -0.111 & -0.092 & -0.120 & 0.056 & 0.060 & -0.103 & -0.129 & -0.117 & -0.135\\
 & (0.092) &  (0.084) & (0.088) & (0.080) & (0.083) & (0.099) & (0.110) & (0.116) & (0.095) & (0.095) & (0.093) & (0.087)\\
\cmidrule(lr){2-6} \cmidrule(lr){7-13}
Number of controls &  & \multicolumn{1}{r}{24} & \multicolumn{1}{r}{32} &  &  &  & \multicolumn{1}{r}{14} & \multicolumn{1}{r}{22} &  &  &  & \\
Sample size &  & \multicolumn{1}{r}{5,839} & \multicolumn{1}{r}{5,839} &  &  &  & \multicolumn{1}{r}{3,938} & \multicolumn{1}{r}{3,938} &  &  &  & 
\\
    \bottomrule
\end{tabular}
\begin{tablenotes}\item\emph{Notes:} This table reports the coefficients for \citet{brownback2020instruction} for all estimators described in Section \ref{sec:solutions}. This table starts by replicating the coefficients from Column 2 in Table 2 of the cited article. Since overlap failed, we report both sets of estimands: full and overlap sample. This article is an experimental study. We report standard errors clustered by instructor in parentheses.
\end{tablenotes}
\end{threeparttable}
\end{table}
\end{landscape}

\newpage

\begin{table}
\begin{threeparttable}
\caption{Full results: \citet{lee2020electrification}, Table 2 - Column 2} \label{tab:art-14}
\begin{tabular}{@{}l @{}cSSSSS SSSSSS @{}}
    \toprule \hline
    & \multicolumn{7}{c}{Full sample} \\
    \cmidrule(lr){2-8}
    Estimates & \multirow{2}{*}{$\hat{\beta}$} & \multicolumn{2}{c}{ATE} & \multicolumn{2}{c}{EW} & \multicolumn{2}{c}{CW}\\
    \cmidrule(lr){3-4} \cmidrule(lr){5-6} \cmidrule(lr){7-8}
    & & {Linear} & {Spline} & {Linear} & {Spline} & {Linear} & {Spline}\\ \hline
    
Low Subsidy & 5.943 & 5.304 & 1.916 & 5.652 & 5.555 & 5.275 & 5.057\\
 & (1.494) & (1.390) & (4.029) & (1.405) & (1.389) & (1.370) & (1.221)\\
Medium Subsidy & 22.882 & 22.580 & 69.688 & 22.880 & 22.604 & 22.252 & 21.519\\
 & (4.003) & (3.948) & (65.278) & (3.984) & (3.937) & (3.835) & (3.420)\\
High Subsidy & 94.972 & 94.114 & 63.558 & 94.594 & 94.677 & 93.986 & 93.984\\
 & (1.268) & (1.197) & (18.790) & (1.193) & (1.140) & (1.245) & (1.215)\\
\hline
Number of controls &  & \multicolumn{1}{r}{14} & \multicolumn{1}{r}{34} &  &  &  & \\
Sample size &  & \multicolumn{1}{r}{2,176} & \multicolumn{1}{r}{2,176} &  &  &  & 
\\
    \bottomrule
\end{tabular}
\begin{tablenotes}\item\emph{Notes:} This table reports the coefficients for \citet{lee2020electrification} for all estimators described in Section \ref{sec:solutions}. This table starts by replicating the coefficients from Column 2 in Table 2 of the cited article. This article is an experimental study. We report standard errors clustered at the community level in parentheses.
\end{tablenotes}
\end{threeparttable}
\end{table}

\newpage

\begin{table}
\begin{threeparttable}
\caption{Full results: \citet{drexler2014financial}, Table 2 - Column 2} \label{tab:drexler}
\begin{tabular}{@{}l @{}cSSSSS SSSSSS @{}}
    \toprule \hline
    & \multicolumn{7}{c}{Full sample} \\
    \cmidrule(lr){2-8}
    Estimates & \multirow{2}{*}{$\hat{\beta}$} & \multicolumn{2}{c}{ATE} & \multicolumn{2}{c}{EW} & \multicolumn{2}{c}{CW}\\
    \cmidrule(lr){3-4} \cmidrule(lr){5-6} \cmidrule(lr){7-8}
    & & {Linear} & {Spline} & {Linear} & {Spline} & {Linear} & {Spline}\\ \hline
    
Standard Accounting & 0.036 & 0.040 & 0.041 & 0.037 & 0.038 & 0.040 & 0.040\\
 & (0.053) & (0.054) & (0.052) & (0.053) & (0.053) & (0.054) & (0.053)\\
Rule-of-Thumb & 0.109 & 0.113 & 0.117 & 0.112 & 0.113 & 0.113 & 0.115\\
 & (0.030) & (0.030) & (0.044) & (0.030) & (0.031) & (0.030) & (0.032)\\
\hline
Number of controls &  & \multicolumn{1}{r}{ 7} & \multicolumn{1}{r}{11} &  &  &  & \\
Sample size &  & \multicolumn{1}{r}{796} & \multicolumn{1}{r}{796} &  &  &  & 
\\
    \bottomrule
\end{tabular}
\begin{tablenotes}\item\emph{Notes:} This table reports the coefficients for \citet{drexler2014financial} for all estimators described in Section \ref{sec:solutions}. This table starts by replicating the coefficients from Column 2 in Table 2 of the cited article. This article is an experimental study. We report heteroskedasticity-robust standard errors in parentheses.
\end{tablenotes}
\end{threeparttable}
\end{table}

\newpage

\begin{landscape}
\begin{table}
\begin{threeparttable}
\caption{Full results: \citet{benhassine2015shove}, Table 5 - Column 1} \label{tab:benhassine}
\setlength{\tabcolsep}{3pt}
\begin{tabular}{@{}l @{}cSSSS cSSSSSS @{}}
    \toprule \hline
    & \multicolumn{5}{c}{Full sample} & \multicolumn{7}{c}{Overlap} \\
    \cmidrule(lr){2-6} \cmidrule(lr){7-13}
    Estimates & \multirow{2}{*}{$\hat{\beta}$} & \multicolumn{2}{c}{EW} & \multicolumn{2}{c}{CW} &
                \multirow{2}{*}{$\hat{\beta}$} & \multicolumn{2}{c}{ATE} & \multicolumn{2}{c}{EW} & \multicolumn{2}{c}{CW} \\
    \cmidrule(lr){3-4} \cmidrule(lr){5-6} \cmidrule(lr){8-9} \cmidrule(lr){10-11} \cmidrule(lr){12-13}
    & & {Linear} & {Spline} & {Linear} & {Spline} &
      & {Linear} & {Spline} & {Linear} & {Spline} & {Linear} & {Spline}\\ \hline
    
LCT to fathers & 0.074 & 0.089 & 0.092 & 0.056 & 0.055 & 0.067 & 0.078 & 0.082 & 0.076 & 0.080 & 0.061 & 0.064\\
 & (0.016) &  (0.017) & (0.016) & (0.018) & (0.018) & (0.019) & (0.015) & (0.015) & (0.020) & (0.020) & (0.020) & (0.020)\\
LCT to mothers & 0.078 &  0.067 & 0.069 & 0.071 & 0.074 & 0.081 & 0.079 & 0.083 & 0.074 & 0.077 & 0.068 & 0.072\\
 & (0.014) &  (0.013) & (0.013) & (0.017) & (0.017) & (0.017) & (0.014) & (0.014) & (0.015) & (0.015) & (0.017) & (0.017)\\
CCTs to fathers & 0.055 &  0.062 & 0.062 & 0.041 & 0.038 & 0.047 & 0.033 & 0.033 & 0.039 & 0.039 & 0.038 & 0.037\\
 & (0.014) &  (0.013) & (0.013) & (0.018) & (0.017) & (0.016) & (0.014) & (0.015) & (0.016) & (0.016) & (0.017) & (0.017)\\
CCTs to mothers & 0.053 & 0.045 & 0.045 & 0.040 & 0.038 & 0.039 & 0.042 & 0.040 & 0.041 & 0.041 & 0.040 & 0.037\\
 & (0.013) &  (0.013) & (0.014) & (0.018) & (0.017) & (0.017) & (0.015) & (0.015) & (0.017) & (0.017) & (0.018) & (0.017)\\
\cmidrule(lr){2-6} \cmidrule(lr){7-13}
Number of controls &  & \multicolumn{1}{r}{57} & \multicolumn{1}{r}{64} &  &  &  & \multicolumn{1}{r}{26} & \multicolumn{1}{r}{33} &  &  &  & \\
Sample size &  & \multicolumn{1}{r}{11,074} & \multicolumn{1}{r}{11,074} &  &  &  & \multicolumn{1}{r}{ 6,996} & \multicolumn{1}{r}{ 6,996} &  &  &  & 
\\
    \bottomrule
\end{tabular}
\begin{tablenotes}\item\emph{Notes:} This table reports the coefficients for \citet{benhassine2015shove} for all estimators described in Section \ref{sec:solutions}. This table starts by replicating the coefficients from Column 1 in Table 5 of the cited article. Since overlap failed, we report both sets of estimands: full and overlap sample. This article is an experimental study. We report standard errors clustered at the school sector level in parentheses.
\end{tablenotes}
\end{threeparttable}
\end{table}

\begin{table}
\begin{threeparttable}
\caption{Full results: \citet{duflo2015education}, Table 2, Panel A - Column 1} \label{tab:duflo}
\setlength{\tabcolsep}{3pt}
\begin{tabular}{@{}l @{}cSSSS cSSSSSS @{}}
    \toprule \hline
    & \multicolumn{5}{c}{Full sample} & \multicolumn{7}{c}{Overlap} \\
    \cmidrule(lr){2-6} \cmidrule(lr){7-13}
    Estimates & \multirow{2}{*}{$\hat{\beta}$} & \multicolumn{2}{c}{EW} & \multicolumn{2}{c}{CW} &
                \multirow{2}{*}{$\hat{\beta}$} & \multicolumn{2}{c}{ATE} & \multicolumn{2}{c}{EW} & \multicolumn{2}{c}{CW} \\
    \cmidrule(lr){3-4} \cmidrule(lr){5-6} \cmidrule(lr){8-9} \cmidrule(lr){10-11} \cmidrule(lr){12-13}
    & & {Linear} & {Spline} & {Linear} & {Spline} &
      & {Linear} & {Spline} & {Linear} & {Spline} & {Linear} & {Spline}\\ \hline
    
Educ.~subsity & -0.031 &  -0.036 & -0.030 & -0.029 & -0.030 & -0.024 & -0.025 & -0.025 & -0.032 & -0.032 & -0.027 & -0.027\\
 & (0.012) &  (0.011) & (0.010) & (0.011) & (0.012) & (0.013) & (0.007) & (0.007) & (0.011) & (0.011) & (0.010) & (0.010)\\
HIV education & 0.003 &  0.009 & 0.001 & 0.002 & 0.002 & -0.000 & 0.003 & 0.001 & 0.005 & 0.002 & -0.000 & -0.004\\
 & (0.011) &  (0.009) & (0.008) & (0.012) & (0.012) & (0.011) & (0.007) & (0.007) & (0.010) & (0.009) & (0.011) & (0.010)\\
Both & -0.016 &  -0.019 & -0.017 & -0.020 & -0.019 & -0.012 & -0.007 & -0.005 & -0.009 & -0.008 & -0.012 & -0.012\\
 & (0.012) &  (0.010) & (0.010) & (0.011) & (0.012) & (0.012) & (0.007) & (0.008) & (0.010) & (0.010) & (0.010) & (0.010)\\
\cmidrule(lr){2-6} \cmidrule(lr){7-13}
Number of controls &  & \multicolumn{1}{r}{86} & \multicolumn{1}{r}{93} &  &  &  & \multicolumn{1}{r}{79} & \multicolumn{1}{r}{82} &  &  &  & \\
Sample size &  & \multicolumn{1}{r}{9,116} & \multicolumn{1}{r}{9,116} &  &  &  & \multicolumn{1}{r}{8,664} & \multicolumn{1}{r}{8,664} &  &  &  & 
\\
    \bottomrule
\end{tabular}
\begin{tablenotes}\item\emph{Notes:} This table reports the coefficients for \citet{duflo2015education} for all estimators described in Section \ref{sec:solutions}. This table starts by replicating the coefficients from Panel A, Column 1 in Table 2 of the cited article. Since overlap failed, we report both sets of estimands: full and overlap sample. This article is an experimental study. We report standard errors clustered at the school level in parentheses.
\end{tablenotes}
\end{threeparttable}
\end{table}
\end{landscape}

\newpage

\begin{landscape}
\begin{table}
\begin{threeparttable}
\caption{Full results: \citet{fryerlevitt2013racial}, Table 3 - Column 4} \label{tab:FL}
\setlength{\tabcolsep}{3pt}
\begin{tabular}{@{}l @{}cSSSS cSSSSSS @{}}
    \toprule
    & \multicolumn{5}{c}{Full sample} & \multicolumn{7}{c}{Overlap} \\
    \cmidrule(lr){2-6} \cmidrule(lr){7-13}
    Estimates & \multirow{2}{*}{$\hat{\beta}$} & \multicolumn{2}{c}{EW} & \multicolumn{2}{c}{CW} &
                \multirow{2}{*}{$\hat{\beta}$} & \multicolumn{2}{c}{ATE} & \multicolumn{2}{c}{EW} & \multicolumn{2}{c}{CW} \\
    \cmidrule(lr){3-4} \cmidrule(lr){5-6} \cmidrule(lr){8-9} \cmidrule(lr){10-11} \cmidrule(lr){12-13}
    & & {Linear} & {Spline} & {Linear} & {Spline} &
      & {Linear} & {Spline} & {Linear} & {Spline} & {Linear} & {Spline}\\ \hline
    
Black & -0.213 &  -0.193 & -0.193 & -0.202 & -0.203 & -0.191 & -0.231 & -0.237 & -0.171 & -0.171 & -0.195 & -0.197\\
 & (0.032) &  (0.034) & (0.034) & (0.065) & (0.066) & (0.037) & (0.038) & (0.038) & (0.040) & (0.040) & (0.059) & (0.059)\\
Hispanic & -0.249 & -0.257 & -0.257 & -0.171 & -0.171 & -0.209 & -0.196 & -0.195 & -0.220 & -0.220 & -0.171 & -0.171\\
 & (0.028) &  (0.030) & (0.030) & (0.046) & (0.046) & (0.032) & (0.033) & (0.033) & (0.034) & (0.034) & (0.045) & (0.045)\\
Asian & -0.294 &  -0.324 & -0.324 & -0.330 & -0.333 & -0.275 & -0.150 & -0.212 & -0.283 & -0.285 & -0.317 & -0.325\\
 & (0.035) &  (0.038) & (0.038) & (0.085) & (0.085) & (0.039) & (0.058) & (0.042) & (0.043) & (0.043) & (0.082) & (0.084)\\
Other & -0.132 &  -0.116 & -0.115 & -0.127 & -0.127 & -0.127 & -0.084 & -0.083 & -0.105 & -0.105 & -0.105 & -0.103\\
 & (0.038) &  (0.039) & (0.039) & (0.046) & (0.046) & (0.043) & (0.035) & (0.035) & (0.044) & (0.044) & (0.047) & (0.048)\\
\cmidrule(lr){2-6} \cmidrule(lr){7-13}
Number of controls &  & \multicolumn{1}{r}{176} & \multicolumn{1}{r}{176} &  &  &  & \multicolumn{1}{r}{127} & \multicolumn{1}{r}{127} &  &  &  & \\
Sample size &  & \multicolumn{1}{r}{8,806} & \multicolumn{1}{r}{8,806} &  &  &  & \multicolumn{1}{r}{6,623} & \multicolumn{1}{r}{6,623} &  &  &  & 
\\
    \bottomrule
\end{tabular}
\begin{tablenotes}\item\emph{Notes:} 
This table reports the coefficients for \citet{fryerlevitt2013racial} for all estimators described in Section \ref{sec:solutions}. This table starts by replicating the coefficients from Column 4 in Table 3 of the cited article. Since overlap failed, we report both sets of estimands: full and overlap sample. This article is an observational study. We report heteroskedasticity-robust standard errors in parentheses. Both methods (with and without splines) have the same number of controls due to the following: we follow \citet{gphk2024contamination} and insert orthogonal polynomials of the continuous variables in the linear regression. As these are fifth-degree polynomials, they add four new covariates, while the spline also introduces four new variables (two extra powers and two knots).
\end{tablenotes}
\end{threeparttable}
\end{table}

\begin{table}
\begin{threeparttable}
\caption{Full results: \citet{rim2020disparities}, Table 2 - Column 3} \label{tab:rim}
\setlength{\tabcolsep}{3pt}
\begin{tabular}{@{}l @{}cSSSS cSSSSSS @{}}
    \toprule
    & \multicolumn{5}{c}{Full sample} & \multicolumn{7}{c}{Overlap} \\
    \cmidrule(lr){2-6} \cmidrule(lr){7-13}
    Estimates & \multirow{2}{*}{$\hat{\beta}$} & \multicolumn{2}{c}{EW} & \multicolumn{2}{c}{CW} &
                \multirow{2}{*}{$\hat{\beta}$} & \multicolumn{2}{c}{ATE} & \multicolumn{2}{c}{EW} & \multicolumn{2}{c}{CW} \\
    \cmidrule(lr){3-4} \cmidrule(lr){5-6} \cmidrule(lr){8-9} \cmidrule(lr){10-11} \cmidrule(lr){12-13}
    & & {Linear} & {Spline} & {Linear} & {Spline} &
      & {Linear} & {Spline} & {Linear} & {Spline} & {Linear} & {Spline}\\ \hline
    
Black & -4.059 &  -3.907 & -3.803 & -3.786 & -3.478 & -4.441 & 8.071 & 3.753 & -3.199 & -3.303 & -3.266 & -3.368\\
 & (1.107) &  (1.210) & (1.168) & (1.597) & (1.541) & (1.149) & (11.922) & (10.928) & (1.039) & (1.119) & (1.403) & (1.407)\\
Hispanic & -1.119 &  -0.837 & -0.841 & 1.290 & 1.515 & -0.658 & 2.927 & 3.366 & -0.879 & -1.043 & -1.099 & -1.179\\
 & (0.731) &  (0.698) & (0.695) & (3.949) & (4.014) & (1.603) & (3.403) & (3.147) & (1.446) & (1.447) & (2.460) & (2.403)\\
Asian & -2.536 &  -2.117 & -2.209 & -4.375 & -4.006 & -3.383 & -8.439 & -19.792 & -3.633 & -3.691 & -3.685 & -3.628\\
 & (0.978) &  (1.206) & (1.193) & (2.896) & (3.007) & (1.440) & (3.606) & (15.363) & (0.930) & (1.055) & (1.824) & (1.861)\\
\cmidrule(lr){2-6} \cmidrule(lr){7-13}
Number of controls &  & \multicolumn{1}{r}{268} & \multicolumn{1}{r}{274} &  &  &  & \multicolumn{1}{r}{ 35} & \multicolumn{1}{r}{ 38} &  &  &  & \\
Sample size &  & \multicolumn{1}{r}{4,037} & \multicolumn{1}{r}{4,037} & &  &  & \multicolumn{1}{r}{  620} & \multicolumn{1}{r}{  620} &  &  &  & 
\\
    \bottomrule
\end{tabular}
\begin{tablenotes}\item\emph{Notes:} This table reports the coefficients for \citet{rim2020disparities} for all estimators described in Section \ref{sec:solutions}. This table starts by replicating the coefficients from Column 3 in Table 2 of the cited article. Since overlap failed, we report both sets of estimands: full and overlap sample. This article is an observational study. We report standard errors clustered at the instructor level in parentheses.
\end{tablenotes}
\end{threeparttable}
\end{table}
\end{landscape}

\end{document}